% fichero tiedin.tex, 23-7-1996, Tresguerres
\magnification=\magstep1
\hsize=13cm
\vsize=20cm
\overfullrule 0pt
\baselineskip=13pt plus1pt minus1pt
%\baselineskip=2\baselineskip
\lineskip=3.5pt plus1pt minus1pt
\lineskiplimit=3.5pt
\parskip=4pt plus1pt minus4pt

% macro for slash
\def\negenspace{\kern-1.1em}

%\def \Behauptung{  
%= \hbox to 0pt{ \kern -10pt \lower 3pt \vbox to 15pt {\hbox     
%{$\scriptstyle ! \> {} $} \vglue 6pt \hbox {} }} }  

%Macro for section, subsection and equation numbers:

\newcount\secno
\secno=0
\newcount\susecno
\newcount\fmno\def\z{\global\advance\fmno by 1 \the\secno.
                       \the\susecno.\the\fmno}
\def\section#1{\global\advance\secno by 1
                \susecno=0 \fmno=0
                \centerline{\bf \the\secno. #1}\par}
\def\subsection#1{\medbreak\global\advance\susecno by 1
                  \fmno=0
       \noindent{\the\secno.\the\susecno. {\it #1}}\noindent}

%Macro for d'Alembertian:

\def\sqr#1#2{{\vcenter{\hrule height.#2pt\hbox{\vrule width.#2pt
height#1pt \kern#1pt \vrule width.#2pt}\hrule height.#2pt}}}

%Macro for footnotes:

\newcount\refno
\refno=1
\def\y{\the\refno}
\def\myfoot#1{\footnote{$^{(\y)}$}{#1}
                 \advance\refno by 1}

%Macro for list of references:

%Macro for not equal:
\def\neq{\hbox{$\,$=\kern-6.5pt /$\,$}}

%Macro for = with * on top:

%Macro for boldface omega (\clom):
%AM font:
%\font\fbg=ambi10\def\clom{\hbox{\fbg\char33}}

%CM font (when using IPS):
%\font\fbg=cmmib10\def\clom{\hbox{\fbg\char33}}

%Macro for section, and equation numbers: (Use \sectio)

\newcount\secno
\secno=0
\newcount\fmno\def\z{\global\advance\fmno by 1 \the\secno.
                       \the\fmno}
\def\sectio#1{\medbreak\global\advance\secno by 1
                  \fmno=0
       \noindent{\the\secno. {\it #1}}\noindent}

%Macro for semidirect product
\def\semidirect{\;{\rlap{$\supset$}\times}\;}

%Pound sterling:= {\it \$\/}50--00.

%To use any special fonts with IPS, include the corresponding
%CM font definition in your macro or TeX file. In most cases this
%coincides with the AM font except for the first letter, e.g. cmr9
%and amr9 (note: use lower case letters). The one major exception is
%the bold symbols font, which is cmmib10 in place of ambi10, etc.

%===================================================================== 
\centerline{\bf{TIME EVOLUTION IN DYNAMICAL SPACETIMES}} 
\vskip 1.0cm 
\centerline{by}
\vskip 1.0cm
\centerline{A. Tiemblo and R. Tresguerres}
\vskip 1.0cm
\centerline{\it {IMAFF, Consejo Superior de Investigaciones Cient\'ificas,}}
\centerline{\it {Serrano 123, Madrid 28006, Spain}}
\vskip 1.5cm
\centerline{ABSTRACT}\bigskip 
We present a gauge--theoretical derivation of the notion of
time, suitable to describe the Hamiltonian time evolution of 
gravitational systems. It is based on a nonlinear coset
realization of the Poincar\'e group, implying the time component
of the coframe to be invariant, and thus to represent a metric
time. The unitary gauge fixing of the boosts gives rise to the 
foliation of spacetime along the time direction. The three 
supressed degrees of freedom correspond to Goldstone--like
fields, whereas the remaining time component is a Higgs--like
boson. 
\bigskip\bigskip 
\sectio{\bf{Introduction}}\bigskip 
The notion of time is fundamental in classical Physics. The dynamical 
laws are understood to be the expressions of the evolution of
any physical system in time. This becomes particularly
evident in the Hamiltonian formulation of mechanics, since 
the Poisson bracket of any phase space variable with the 
Hamiltonian yields the time derivative. Further, Dirac's
standard quantization procedure by means of the correspondence 
principle rests on the Hamiltonian formalism. Thus, the
requirement of disposing of a reasonable characterization of
time is necessary for both, the classical and the quantum
dynamical approaches. However, serious problems arise when one 
tries to formulate Gravitation as a Hamiltonian theory$^{(1)}$ 
--as a first step to quantize it-- due to the difficulty in 
defining a suitable general--relativistic time. Our main task
will be to identify such a time with respect to which the 
Hamiltonian evolution of a gravitational system makes sense. 

Which are the minimal features one should require from time?
Rovelli$^{(1)}$ has classified the main time properties as they
manifest themselves in different physical theories. Since they
are mutually contradictory, it becomes manifest that the notion
of time is not an unified one in Physics. But at least, we would
like to have a criterion to judge which time features are more 
desirable to be mantained in the context of General Relativity. 
Several ones are characteristic for Newtonian mechanics and are 
lost in relativistic Physics. We can renounce to them without
any trouble, notwithstanding the fact that we can rencounter
them in particular circumstances. We mean mainly the uniqueness
and the spatial globallity, i.e. the possibility of measuring 
the same preferred time variable in all spatial points. Another
characteristic of time, which is common to Newtonian and
special--relativistic time, is to be external. In other words, it
is suposed to exist separately, with independence of the
dynamics. This point is particularly interesting. As we will see
in the following, in the context of gravitational theories the
{\it status} of time is different. When suitably identified, the 
time becomes internal, i.e. dynamically determined. Although
this may constitute a source of difficulties, if one could solve
them, then one had identified, from Gravity, the dynamical origin 
of the time variable appearing as external in any other physical
theory.

The internal nature of time is thus an appealing feature we will
have to deal with. However, there are other aspects of time
which are more evident, and they should also be present in the
physical time we are looking for. First of all, one expects the
time to be one--dimensional. Further, it should be metrical, in
order to make it possible to compare distinct time intervals to
each other, and temporally global, i.e. such that every event 
goes through any value of the time variable once and only once. 
In addition to the sketched features that one expects from time as
far as Gravitation is involved, the resulting time notion should
be suitable to construct a consistent Hamiltonian formulation of
the gravitational theory. Let us now look at the present situation
in the development of this program. 

In Newtonian mechanics, the topology of space and time is given
by the Cartesian product $E^3\times R$ of the three--dimensional
Euclidean space times the real line representing the time. The
time intervals are invariant under Galilean transformations, and
thus unique for any two events. Contrarily, the uniqueness of
time is absent from relativistic Physics. In fact, in Special 
Relativity there exist a three parameter family of times, 
depending on the relative three--velocity, representable as lines
filling the light cone. 

The role played by time in General Relativity is considerably
more confusing, due to the fact that the spacetime manifold is
treated as a whole. As far as the field equations retain their
original four--dimensional form, no problem seems to arise. But 
the general covariance of the theory avoids to identify a well
defined separated time notion. In principle, only locally can 
one establish a soldering to the tangent spaces, locally 
representing inertial frames, in which the distinction between
spatial and temporal directions reduces to that of Special 
Relativity. The time would then only posses a local meaning. 
Without an important additional assumption, the ideas of a 
cosmological time, of the age of the universe etc., i.e. all 
conceptions of time which presupose its global nature, would not
make any sense. The cosmological time normally used in Cosmology
arises from particular solutions of the Einstein equations, like
the Friedmann--Robertson--Walker solution, in which the alluded 
assumption is implicitely included. We mean the following. From
a general theoretical point of view, the only way to recover a 
global time notion in the context of General Relativity, similar
in some extent to the classical or special--relativistic ones,
is to perform a foliation of the four--dimensional spacetime
manifold along a certain direction identifyable as the time
direction itself. In order to do so, one has to impose the
Frobenius foliation condition. It can be introduced in different
manners, each of which leading to a particular time notion. We
will analize it in detail in the following. 

For the moment, let us briefly review the three standard 
approaches to time in General Relativity$^{(1)}$. All of them 
present theoretical difficulties which make them unsatisfactory
as expressions of a well defined notion of physical time. In 
particular, Rovelli points out that none of them is applyable to
a rigorous quantum treatment of Gravity. But their faults
manifest themselves already at the classical level, as we will
see immediately. 

In the first place, the most naive attempt is to identify the
time coordinate at any point as the general--relativistic time. 
This is the so called coordinate time. We mentioned above that 
such a time is, by its own nature, local. Further, in view of 
the general covariance of General Relativity, the coordinate
time consists of an infinite--dimensional family of time lines
which are arbitrarily rescalable. Accordingly, such a time 
lacks on time metricity. The difference between two values of
the time parameter is not really interpretable as a time
interval, since no time metric is defined which guarantees this 
extent. On the other hand, when considered from the point of
view of the gauge approach to Gravity$^{(2)}$, it also lacks on 
any symmetry under gauge transformations. We conclude that the 
coordinate time is to be disregarded as the true physical time. 

A better candidate is the proper time along any time--like
worldline. It has the advantage over the coordinate time of
being metrical, thus allowing to determine which time intervals
are of equal duration. In fact, it is invariant under arbitrary 
coordinate transformations. It is also time global in the sense 
that every event --in this case every solution of the field 
equations-- {\it passes} through every value of the time
parameter once an only once. In contrast to all
pre--relativistic approaches to the notion of time, it is not 
external to the theory, i.e. it is not given {\it a priori} but 
dynamically determined, since it depends on the whole $4\times 4$ 
metric. A serious difficulty arises in the understanding of
dynamics as evolution in this time variable which is itself 
subordined to the dynamics. It is the following. We are
confronted with the paradoxical fact that one has to formulate
and solve the field equations in terms of the coordinate time,
which has no metric properties, because proper time is still not
at our disposal. Only after having solved the dynamical problem
can proper time be defined. Consequently, the gravitational
dynamics cannot be expressed as evolution with respect to the
proper time variable possessing the right metric properties. 

Let us look at the standard solution to this problem. It is
given by a third kind of general--relativistic time, 
namely what is called the {\it clock time}. Rovelli presentes it as
"the only way of recovering a conventional Hamiltonian evolution in
General Relativity". In fact, it constitutes an alternative to 
the previous dylemma between time as evolution parameter without 
metric properties (the coordinate time), and metric time (the 
proper time), not present at the level of the evolution
equations. The clock time is the time measured by a physical 
clock, or more exactly, measured with respect to a dynamical 
variable, chosen as a clock. It allows to express the evolution
of the whole system relatively to a time with metric properties. 
One has to consider General Relativity coupled to matter, and 
express the gravitational variables as functions of the matter 
ones. The role of physical time is then played by one of the 
(gauge invariant) degrees of freedom of the theory itself, 
standing for the {\it clock}. Being the latter a physical
object, its rhythm depends on the field equations themselves.
Thus it is an internal time. An example of clock time is
provided by the radius of the universe, used as time variable in
Cosmology. It is metrical; however, the temporal globality is 
absent from this clock time if the universe collapses. But the 
main difficulty with clock time is a more fundamental one. It
has to do with its possible application to a Hamiltonian
treatment. The clock time allows to recover a well defined time 
evolution in General Relativity, but only a parametrized
evolution, not a genuine one. Unfortunately, parametrized 
Hamiltonian systems are not well understood as quantum systems. 

The aim of the present paper is to give an answer to these 
troubles. We will develope an alternative notion of dynamical
time which possesses the main features necessary to solve the 
problem of defining a standard Hamiltonian time evolution of
Gravitation, and of any other system in the presence of Gravity.
Further, if the gravitational effects are put off, the time to
be introduced below remains as a consistent definition of time.
It is an invariant time entering in a natural way the dynamical
field equations. In some extent, it may be understood as a 
preferred clock time. In fact, it is a dynamical field of the 
theory, whose value is affected by the matter sources. The
evolution of any physical system is evaluated with respect to
it. However, it is not an arbitrarily chosen physical {\it
clock}, but the time component of the coframe itself, suitably
constructed to be Poincar\'e gauge invariant. Being invariant, it 
guarantees the time metricity, i.e. the comparability of time 
intervals. In the unitary gauge to be studied in detail in the 
following, the time field takes the form $\vartheta ^0 =
u^0\,d\tau\,$, inducing a foliation of the spacetime along it.   

The key to realize the program of constructing a complete
characterization of the physical spacetime, including a time
with the desired properties, is provided by a particular 
nonlinear realizations of a certain spacetime symmetry group, 
in particular the Poincar\'e group, acting on its own parameter 
space. The abstract group will be our unique departing point. 
We will not need to postulate additional mathematical structures
to fulfil our scheme, such as a pre--dynamical spacetime 
manifold providing the coordinates, or Cartan's {\it rep\`eres 
mobiles} representing the reference frames. No element exterior 
to the gauge group will be present. From the basic assumption
that the Poincar\'e group is a fundamental physical symmetry, we 
will be able to derive simultaneously both, the differentiable 
coordinate manifold defining the topology of spacetime, and the 
dynamical fields attached to it, standing in particular for the 
coframes, curvature, etc. 

The paper is organized as follows. In section 2, we briefly
summarize the nonlinear coset realization procedure which 
constitutes the mathematical basis of the present work.
Section 3 is devoted to the discussion of the origin of
coordinates in the nonlinear approach, and some considerations
are made about their physical meaning. In section 4, we apply
the previous general results to the Poincar\'e group, which is
realized nonlinearly in a particular way. As a result, a 
Poincar\'e invariant time component $\vartheta ^0$ of the coframe
arises, which is proposed as the candidate to play the role of 
the physical time. Further, in section 5 we discuss the Poincar\'e
invariant spacetime foliation along $\vartheta ^0$. After 
reviewing the main features of the unitary gauge in section 6, 
we devote section 7 to apply it in the context of Gravity, in
such a way that we derive the previously studied invariant foliation
condition as a result of suitably covariantly fixing the gauge.
\bigskip\bigskip 

\sectio{\bf Nonlinear coset realizations}
\bigskip 
The nonlinear coset approach was originally introduced by 
Coleman et al.$^{(3)}$ in the context of internal symmetry
groups. It was soon extended to spacetime symmetries$^{(4)}$,
and we have shown in several previous papers$^{(5,6)}$ that it 
constitutes the natural framework to construct gauge theories of
Gravity founded on different spacetime groups. The nonlinear 
realizations allow to define the coframes in terms of gauge
fields. They are identified as the nonlinear connections of the 
translations$^{(5,7)}$. The metric tensor does not play any
dynamical role since the gravitatinal forces are carried
exclusively by nonlinear gauge fields. Recently, we 
have proposed a Hamiltonian treatment of the Poincar\'e Gauge
Theory of Gravitation$^{(6)}$ based on a particular nonlinear 
realization of the Poincar\'e group, and we were able to derive
the Einstein equations and the complete set of constraints of
the theory, giving account of the Ashtekar approach. Since the 
departing point to all these results is constituted by the 
nonlinear realizations, here we will outline their essential 
features.

Let $G=\{g\}$ be a Lie group including a subgroup $H=\{h\}$ 
whose linear representations $\rho (h)$ are known, acting on 
functions $\psi $ belonging to a linear representation space of 
$H$. We distinguish between $G$ considered as a transformation 
group, and the group $G$ itself as a differentiable manifold. 
In order to define the nonlinear action of $G$ on its own
group manifold, we characterize the latter as a principal fibre
bundle $G(M\,,H\,)$ with base space $M=G/H$ and structure 
group $H$ as follows$^{(8)}$.

Let the subgroup $H$ act freely on $G$ on the right, i.e. 
$\forall\,h\,\epsilon\,H\,,\forall\,g\,\epsilon\,G\,,\,R_h g 
:=gh\,.$ This action induces an equivalence relation between 
elements $g\,,g\,'\,\epsilon\,G\,$, defined as
$$g\,'\equiv g\Leftrightarrow\exists\,h\,\epsilon\,H\,/\,g\,'= 
R_h g\,,\eqno(\z)$$
which gives rise to a complete partition of the group manifold 
$G$ into equivalence classes $g\,H$, namely
$$g\,H:=\left\{ R_h g\,/\,g\,\epsilon\,G\,,\forall\,h\,
\epsilon\,H\,\right\}\,.\eqno(\z)$$
The quotient space $G/H$ of $G$ by the equivalence relation
induced by $H$ is taken to be the base manifold of the fibre 
bundle. Its elements are single representatives of each
equivalence class. Since we deal with Lie groups, the elements
of $G/H$ are characterized by continuous coset parameters, say
$\xi\,$, playing the role of coordinates. We identify the 
canonical projection $\pi : G\rightarrow G/H$ of the fibre
bundle to be the mapping from equivalent points $g$ and $R_h g$ 
to the same point $\xi\,\epsilon\, G/H\,$. The equivalence class
$\pi ^{-1}\left(\xi\,\right) = g\,H\,$ of left cosets labeled by
$\xi$ is called the fibre through $g\,$, and it is isomorphic to
the structure group $H\,$. Since the latter is a subgroup of $G$
defined as
$$H:=\{ h\,\epsilon\,G\,/\,\pi\left( R_h\,g\,\right) 
=\pi\left( g\,\right)\,,\,\forall\,g\,\epsilon\,G\,\}\,,
\eqno(\z)$$ 
alternative choices of the canonical projection allow to
structurate the group manifold in different manners, each of
which corresponding to a distinct structure group.

In brief, the group $G$ itself, considered as a manifold, is a 
differentiable principal fibre bundle $G\left( G/H\,,H\right)$
over the base manifold $G/H$ with structure group $H\,.$ 
The fibre bundle has locally the topology of a direct product of
the base space $G/H$ and the fibre $H$, in the sense that
every point $\xi\,\epsilon\,G/H$ has a neighborhood $U$ such
that $\pi ^{-1}\left(U\,\right)$ is isomorphic with $U\times H\,$, 
i.e. there exist a diffeomorphism $\chi : \pi ^{-1}\left(U\,\right)
\rightarrow U\times H$ such that $\chi\left(g\,\right) =
\left( \pi\left(g\,\right)\,,\varphi\left(g\,\right)\right)\,$, where 
$\varphi$ satisfies the condition that $\varphi\left( R_h g\,\right) 
= R_h\varphi\left( g\,\right)\,$. 

Let us now consider $G$ as a transformation group with elements 
parametrized as $g_t\,$. The left action 
$$L_{g_t}\,g:=g_t\,g\eqno(\z)$$
of $g_t\,\epsilon\, G$ on elements $g$ of the group manifold
$G\left( G/H\,,H\right)$ constitutes the basis of the nonlinear
coset realizations of the group. Let the orbit $\xi _t = L_{g_t}
\xi _0$ be a curve through $\xi _0$ on $G/H\,,$ and 
$\pi ^{-1}\left( \xi _t\,\right)$ the fibres over $\xi _t\,.$ We
suppose that the projection $\pi$ is an intertwining map for the
action of $g_t\,\epsilon\,G$ on $G$ and $G/H\,,$ i.e.
$$L_{g_t}\left(\pi ^{-1}\left( \xi _0\,\right)\right) 
=\pi ^{-1}\left( \xi _t\,\right)\,\,.\eqno(\z)$$
Thus, the action of $G$ on the left moves from the fibre over
$\xi _0$ to the fibre over $\xi _t\,.$ Let us consider a family of
sections $\{\sigma\left( \xi\,\right)\}\subset\pi ^{-1}\left(
\xi\,\right)$ whose values on a given fibre are related by $H$
as $\sigma\,'\left( \xi _t\,\right) =R_h\sigma\left(
\xi _t\,\right)\,.$ The action of $G$ will in general change from a
section to another, which is compatible with (2.5). Accordingly, we
can decompose the total left action of $G$ on $\sigma\left(
\xi _0\,\right)$ into a displacement along the section 
$\sigma\left( \xi\,\right)$ from $\sigma\left( \xi _0\,\right)$
to $\sigma\left( \xi _t\,\right)$ followed by a change along the
fibre $\pi ^{-1}\left( \xi _t\,\right)$ from $\sigma
\left( \xi _t\, \right)$ to
the new section $\sigma\,'\left( \xi _t\,\right) = 
R_{h\left( \xi _t\right)}\sigma\left( \xi _t\,\right)\,$, i.e.
$$L_{g_t}\sigma\left( \xi _0\,\right)= R_{h\left( g_t\,,\xi _0
\,\right)}\sigma\left( \xi _t\,\right)\,\,.\eqno(\z)$$
In the following, we will also call the structure group $H$ the 
{\it{classification subgroup}} in order to mantain the
terminology of previous papers$^{(5,6)}$. The fundamental theorem
on nonlinear realizations, due to Coleman et al.,$^{(3)}$
establishes that the elements $g_t$ of the whole group $G$ 
considered in (2.6) act nonlinearly on the representation spaces of 
the classification subgroup $H$ according to
$$\psi '=\,\rho\left(h\left( g_t\,,\xi _0\,\right)\right)\psi\,,
\eqno(\z)$$
where $\rho$, as mentioned above, is a linear representation
of $H$ on the $\psi$ fields. Therefore, the action of the total 
group $G$ projects on the representations of the subgroup $H$
through the dependence of $h\left(g_t\,,\xi _0\,\right)$ in (2.7)
on the group element $g_t$, as given by eq.(2.6). The group is 
realized on the couples $\left(\xi\,,\psi\right)\,$, and it 
reduces to the standard linear action for $H=G\,$.

The usual situation in Physics is that an independent spacetime 
differentiable manifold previous to the dynamics is postulated
to exist. The remaining physical objects, including the
geometrical post--topological structures, are constructed on it.
In contrast, the nonlinear approach allows to derive everything,
including the coordinate base manifold, from the symmetry group.
This will become apparent later, when we deal with spacetime
groups. The coordinates are associated to the translations, and
they appear as parameters of the base space $G/H\,$, as far as a
spacetime group $G$ including translations is taken to be the
gauge group of Gravity. Thus, even the coordinate manifold is 
dynamically derivable in the nonlinear gauge approach to
spacetime. Moreover, being the coordinates group parameters, the
spacetime manifold is automatically differentiable. 

The dynamical content of the physical spacetime, i.e. its 
post--topological structure, is provided by the connections, 
playing the role of gauge fields of a certain (spacetime) group.
Our next step will be to introduce them in the nonlinear scheme.
In terms of a suitable nonlinear connection $\Gamma\,$, it will
be possible to define a covariant differential transforming like
(2.7) under the local action of $G\,$. In order to facilitate 
calculations, let us rewrite (2.6) in the more explicit form 
$$g\,\sigma\,(\xi\,)=\,\sigma\,(\xi ')\, h\left( g\,,\xi\right)
\,,\eqno(\z)$$
with $g$ standing for $g_t\,$, and being $h\left( g\,,\xi
\right)$ the elements of the classification subgroup $H$. 
The nonlinear connection relates to the ordinary linear one 
$\Omega$ as
$$\Gamma =\,\sigma^{-1}\left(d\,+\Omega\,\right)\sigma\,.
\eqno(\z)$$
Since the linear connection $\Omega$ transforms as
$$\Omega '=\, g\,\Omega \,g^{-1}+g\,d\,g^{-1}\,,\eqno(\z)$$
it is easy to check, making use of (2.8), that the nonlinear
connection $\Gamma$ transforms as
$$\Gamma '=\,h \Gamma h^{-1} +h d\, h^{-1}\,.\eqno(\z)$$ 
The nonlinear covariant differential operator constructed in
terms of (2.9) reads 
$${\bf D}:=\,d\, +\Gamma\,.\eqno(\z)$$
>From (2.11) follows that only the components of $\Gamma$ involving 
the generators of $H$ behave as true connections, transforming 
inhomogeneously, whereas the remaining components transform as 
tensors with respect to the subgroup $H\,$, despite their nature
of connections.
\bigskip\bigskip 

\sectio{\bf Some comments on the origin and physical meaning of
coordinates}\bigskip 
Topologically considered, the totality of events in space and 
time constitutes a four--dimensional manifold, since everything 
happens in a place at a certain instant. This fact was firstly 
pointed out by Minkowski$^{(9)}$ in his early deduction of
Special Relativity, alternative to that of Einstein. According
to his terminology, we call any actual or possible event a 
worldpoint, and it is further assumed that the set of all worldpoints
constitutes a differentiable manifold, so that we can assign
locally to any worldpoint four coordinates $(\,x^0,x^1,x^2,x^3\,)$ 
of a suitable chart of $R^4\,$. 
The real coordinates characterize the continuity of the world,
i.e. they describe a topological feature of the arena underlying
any physical event. However, only under very particular
assumptions do they become directly related to observable
quantities. For instance, Cartesian coordinates posses in fact an 
immediate metrical meaning in classical and special relativistic
mechanics. The three spatial (Cartesian) coordinates represent 
lengths identical with the Euclidean projections of a given
event on rigid axes defined overall the space, whereas the time 
coordinate is measured by a clock at rest with respect to the 
spatial axes. Nevertheless, this is far from being the natural 
interpretation of coordinates in general. Moreover, the
observable meaning of coordinates in classical mechanics and
in Special Relativity is the result of a further theoretical 
development consisting in having introduced a rigid metric
tensor in such a way that certain dynamical quantities, namely
the coframes, become expressable directly in terms of the 
coordinates, see below. 

As we have mentioned above, the nonlinear framework, when
applied to a spacetime group including translations, suffices to
yield a differentiable coordinate manifold, thus being
unnecessary to postulate it {\it a priori}, separately from the
spacetime dynamics. To illustrate this point, let us consider the
simple example of the affine group $A(4\,,R)=\, GL(4\,,R)
\semidirect R^{4}$, which is the semidirect product of the
general linear transformations and the translations. Its
respective generators $\Lambda ^\alpha {}_\beta$ and $P_\alpha$ 
satisfy the commutation relations 
$$\left[\Lambda ^\alpha {}_\beta\,,\Lambda ^\mu {}_\nu\right]=
\,i\,\left(\delta^\alpha _\nu \Lambda ^\mu {}_\beta   
         -\delta^\mu _\beta \Lambda ^\alpha {}_\nu\right)\,, 
\quad \left[\Lambda ^\alpha {}_\beta\,, P_\mu\right]=
\,i\,\delta^\alpha _\mu P_\beta\,,
\quad \left[P_\alpha\,, P_\beta\right]=\,0\,.\eqno(\z)$$
The (infinitesimal) group elements of the whole affine group 
$A(4\,,R)$ are parametrized as  
$$g=\,  e^{i\,\epsilon ^{\alpha} P_\alpha}
e^{i\,\zeta _\alpha {}^\beta \Lambda ^\alpha {}_\beta}
\approx 1+i\,\epsilon ^{\alpha} P_\alpha +
i\,\zeta _\alpha {}^\beta \Lambda ^\alpha {}_\beta\,.\eqno(\z)$$
We will realize the group action taking $H=GL(4\,,R)$ as the
structure group. Accordingly, we choose its (infinitesimal) 
elements in (2.8) to be parametrized as 
$$h =e^{i\,u_\alpha {}^\beta \Lambda ^\alpha {}_\beta}
\approx 1+i\,u_\alpha {}^\beta \Lambda ^\alpha {}_\beta
\,.\eqno(\z)$$
The sections  
$$\sigma =e^{-i\,x ^\alpha P_\alpha}\eqno(\z)$$
depend on (finite) parameters $x ^\alpha$ of the base space 
$A(4\,,R)/GL(4\,,R)$. Making use of the fundamental eq.(2.6), or 
equivalently of (2.8), after a little algebra we find the variation
of $x ^\alpha$ to be 
$$\delta\,x ^{\alpha }=-\zeta _\beta {}^\alpha
\,x ^\beta -\epsilon ^\alpha\quad\,,\qquad u _\alpha {}^\beta 
=\,\zeta _\alpha {}^\beta \,.\eqno(\z)$$
This shows that the parameters $x ^\alpha$ associated to
the translations behave in fact as coordinates. Although
presented here in a particular example, the result is general. 

What is the physical {\it status} one should adscribe to the
underlying worldpoint manifold? Do it represent a certain
ontological background --namely the spacetime-- on which the 
events actually take place, or is it to be merely considered 
as a mathematical artefact? This question confronts us with
the old Clarke--Leibniz disputation$^{(10)}$, which may be 
translated into mathematical terms as follows. 

First of all, we have to distinguish between the passive and the
active interpretations of general coordinate transformations. A 
transformation is called passive if it represents a change from
a local coordinate chart to another, whereas the described point
$p$ of the manifold remains the same. Such a change is purely
nominal. On the other hand, an active transformation or
diffeomorphism is a bijective $C^{\infty}$ application between
distinct open sets of the manifold, thus really {\it moving}
from a point to a different one. Let us now see how both sorts
of transformations are interpreted from the absolutistic and the
relationalistic points of view respectively. 

According to the absolute space (resp. spacetime) conception 
defended by Clarke, the topological worldpoint manifold
constitutes a sort of actual receptacle on which the events are 
immerged. The points actually exist as the ultimate constituents 
of the space. Thus, the dependence of the physical variables on 
them shows that the events are actually attached to spacetime 
points. Accordingly, the active coordinate transformations are 
viewed as essentially different from the passive ones. Although
a diffeomorphism (in a diff--invariant formulation of any
physical theory) preserves the reciprocal relations between
events, so that no observable consequences arise, an absolutist 
would notwithstanding recognize the original and the transformed
states as two actually different although empirically 
undistinguishable events, since they are distinctly located on
the absolute spacetime manifold.

The relationalistic viewpoint represented by Leibniz rejects
this interpretation. The points are necessary to represent the 
relative localizations of physical events, but neither are the 
events in fact attached to particular points, nor do the points 
really exist. The points behave as mathematical labels which 
allow to express spatial relations between distinct physical 
variables. Nevertheless, the relations described in this way 
are independent from any actual points. The spatial relations
and not the points of the manifold are the ultimate spatial
reality. In fact, as accepted even by the absolutists, a 
diffeomorphism actively transforming the points involved in 
the description of a (diff--invariant) physical system leaves it
undistinguishable from the original one. According to Leibniz's 
principle of the identity of undiscernibles, both systems are 
identical. Thus, from the relational point of view, the 
empirically irrelevant diff--transformations are fictitious, 
{\it agendo nihil agere}. Not only nothing observable occurs; in
fact nothing occurs absolutely. Since the spacetime points do
not exist actually, whenever the relative configuration of the 
physical objects is not altered, the transformations from a
point to another are viewed as a sort of renaming. Although
matematically different from the passive coordinate
transformations, the diffeomorphisms are also considered by the
relationalists as nominal changes in the sense that they {\it
move} the physical objects through a fictitious (purely 
mathematical) background, not altering their real physical 
features. 

Of course, the relationalistic interpretation is not directly
readable out from the mathematical formulation. But here we are 
not discussing about the mathematical sintaxis of the theory 
but about its semantics. In this sense, the absolutistic point 
of view is closer to the litterality of the language employed to
describe the spatial relations. In fact, it constitutes a naive 
realistic interpretation of the formalism. Everything which has 
a name (in this case the points) is suposed to exist actually, 
despite its non observable nature. However, since we are
concerned with Physics, we can forget the metaphysical belief in
real points. We merely identify them as nominal labels whose
role is that of functional arguments necessary to express
relations, and we recognize the relations themselves as the only
physical reality. Our nominalistic choice has the advantage of 
supressing non observable theoretical structures in order to
deal exclusively with physically relevant objects. But at last,
since the formalism (i.e. its sintaxis) and its possible
predictions remain unaltered by both, the absolutistic and the 
relationalistic semantical interpretations, the reader is free
to choose between them according to his particular taste. 

The previous discussion concerns the local topology. When globally 
considered, the worldpoint manifold is meaningful as the domain of 
possible events. It tells, for instance, about the open or
closed character of the universe, or about the linearity or --let us 
say-- ciclicity of time. However, that is compatible with the fact 
that the points are not real by themselves. The physical
irrelevance of the coordinates invites to use Cartan's intrinsic
formulation$^{(11)}$, in which the geometrical inasmuch as the
nongeometrical physical objects are represented in a coordinate 
independent manner. When expressed in the language of exterior 
calculus, any physical theory manifests itself as invariant
under general coordinate transformations. Consequently, the
differential forms are the natural mathematical objects to 
represent physical quantities without explicit reference to any 
attachement to the {\it underlying} manifold. We will make an 
extensive use of the intrinsic formulation in the following.
\bigskip\bigskip 

\sectio{\bf Nonlinear realizations of the Poincar\'e group, and 
invariant time}
\bigskip 
The physical group symmetries provide a criterion of
objectivity. The objective reality is suposed to be represented 
by group invariants, since they are not affected by
transformations leading from a reference frame to another. The 
covariant laws represent distinct perspectives on the (objective)
group invariants. Thus, although the numerical results of the 
measurements depend on the reference frame, all of them describe
a common reality. 
The choice of the suitable group is essential for any physical
theory. In particular, in the dynamical approach to spacetime, the 
transformations under a suitable symmetry group will desribe the
true --i.e. relative-- motions. Since the physical laws are 
covariant under such transformations, they describe the relative
behavior of physical quantities without specifying any
particular point of view. In this sense, those states which
differ on an active transformation of the gauge group are 
physically equivalent. 

In our approach to the dynamical theory of spacetime we will
choose the Poincar\'e group as the physical spacetime symmetry. We
do so for different reasons. We could appeal to Einstein's 
sincronization principle relating local times, which requires to
dispose of an invariant (objective) ligth velocity, common to
all reference frames. The local validity of Special Relativity 
would then naturally lead to adopt the Poincar\'e group. But the
main argument is the following. In the spirit of gauge theories,
the gauge fields are derived from the local realization of the 
symmetry group of the sources. In particular, since the matter 
should determine the features of the spacetime to which it
couples, it is natural to depart from the Poincar\'e group, since
it is the classification group of the elementary particles. 
Certainly, we observe that this is not the only possible choice.
In fact, we have shown$^{(5)}$ that one can conciliate the
existence of fermionic matter with the gauge theory of more
general spacetime groups including the Poincar\'e group as a
subgroup. But the simplest symmetry group with this feature is 
the Poincar\'e group itself. Thus, we choose it for simplicity, 
although the generalization to other groups, such as the affine 
group$^{(12)}$, remains an open possibility which does not change 
the general result of the present paper on the interpretation of
time.

The abstract Poincar\'e group $P$ has the Minkowski metric 
$o_{\alpha\beta}$ as its natural invariant, i.e. 
$\delta o_{\alpha\beta}=\,0\,$. In addition to the choice of the
symmetry group, it is important the way in which it is realized.
Making use of the nonlinear procedure of section 2, the action
of the group will be defined on its own parameter space. 
Accordingly, the spacetime manifold is provided by the Poincar\'e 
group $P$ itself, considered as a differentiable principal fibre
bundle $P\left( M\,,H\right)$ over the base manifold $M=P/SO(3)$
with structure group $H=SO(3)\,$. This choice allows to single
out the role of a Poincar\'e invariant time. In a single expression, 
the spacetime is represented by the mathematical object
$$P\left( P/SO(3)\,,SO(3)\right)\,,\eqno(\z)$$
on which a nonlinear action of the Poincar\'e group is defined. 
Taking the connections into account, not only the topology, but
the whole post--topological structure of spacetime is
determined dynamically by the abstract gauge group. The
nonlinear translational connections $\vartheta ^\alpha$ are
responsable for the existence of coframes (being their dual
vectors $e_\alpha$ the reference frames), and the field strength
$R_\alpha {}^\beta$ of the Lorentz connections 
$\Gamma _\alpha {}^\beta$ stands for the curvature, all of them 
depending on the material sources. Thus, the dynamics causes all
what is physically observable about spacetime. It precedes even 
the topology and the kinematics. Nor the underlying manifold,
neither the reference frames are given {\it a priori},
previously to their dynamical definition from the Poincar\'e gauge
group. The spacetime geometry is the natural interpretation of
the gauge dynamics of a certain spacetime symmetry group. Since 
the theory of Gravity defines the dynamical spacetime, it
provides the geometrical scenario for the remaining
interactions. With respect to them --for instance in the context
of electrodynamics, or in the standard model--, spacetime
appears as externally given; but this is a consequence of having
taken it from Gravitation, which explains its origin also in
the absence of gravitational forces. The gauge theory of Gravity
is the dynamical theory of spacetime. Relative to it, spacetime 
is necessarily internal, i.e. determined by field equations. 

Let us now derive the main features of spacetime from the 
nonlinear gauge approach to the Poincar\'e group $P\,$. Its Lorentz 
generators $L_{\alpha\beta}$ and the translational generators 
$P_\alpha\,$ $(\alpha\,,\beta =\,0,...3\,)$, satisfy the
usual commutation relations as given in (A.1). In order to clarify 
the role played by the coframes in the nonlinear treatment of
the Poincar\'e group, we will proceed in two steps. First we 
consider some aspects of the nonlinear theory with the Lorentz 
group as the structure group, and then we develop the theory we 
are here interested in, namely that with structure group 
$H=SO(3)\,$. We do so because a simple relation between both 
realizations exists, which helps to understand the nature of the
Poincar\'e invariance of time manifesting itself in the latter 
approach. Briefly, for $H=$Lorentz we choose 
$$g=\,e^{i\,\epsilon^\alpha P_\alpha }
e^{i\,\beta ^{\alpha\beta}L_{\alpha\beta}}\,,\quad 
\tilde{h}=e^{i\,u^{\alpha\beta}L_{\alpha\beta}}\,,\quad 
\tilde{\sigma}=e^{-i\,x ^\alpha P_\alpha}\,,\eqno(\z)$$
to be substituted in (2.8). The tildes are introduced for later
convenience. The nonlinear action yields the coordinate
transformation  
$$\delta x^\alpha =-\beta _\beta {}^\alpha\,x^\beta 
-\epsilon ^\alpha\quad\,,\qquad u^{\alpha\beta}=
\beta ^{\alpha\beta}\,.\eqno(\z)$$ 
The ordinary linear Poincar\'e connection $\Omega$ in (2.9), with 
values on the Lie algebra, reads   
$$\Omega :=-i\,{\buildrel (T)\over{\Gamma ^\alpha}} P_\alpha 
           -i\,\Omega ^{\alpha\beta}L_{\alpha\beta}\,,\eqno(\z)$$
defined on the base space $P/SO(3)$ of coordinate parameters 
$x^\alpha$ associated to the translations.
It includes the translational and the Lorentz contributions 
${\buildrel (T)\over{\Gamma ^\alpha}}$ and $\Omega
^{\alpha\beta}$ respectively. In terms of (4.4), the nonlinear 
connection (2.9) reads
$$\tilde{\Gamma}:=\,\tilde{\sigma}^{-1}\left(d\,+\Omega\,\right)
\tilde{\sigma} =-i\,\tilde{\vartheta}^\alpha P_\alpha 
-i\,\tilde{\Gamma}^{\alpha\beta}L_{\alpha\beta}\,.\eqno(\z)$$
The translational nonlinear connections $\tilde{\vartheta}^\alpha$ 
in (4.5) are to be identified as the 1--form basis geometrically 
interpretable as the coframe$^{(5-7)}$. From (4.2,4,5) we find 
$$\tilde{\vartheta }^\alpha :=\,{\buildrel (T)\over
{\Gamma ^\alpha }}+D\,x^a\quad\,,\qquad
\tilde{\Gamma}^{\alpha\beta}=\Omega ^{\alpha\beta}\,.\eqno(\z)$$
According to (2.11), whereas $\tilde{\Gamma}^{\alpha\beta}$ in (4.5)
remains a true connection, the coframe $\tilde{\vartheta }^\alpha$ 
behaves as a Lorentz four--vector under local Poincar\'e 
transformations.

With these results at hand, we now proceed to realize the
Poincar\'e group nonlinearly with its subgroup $H=SO(3)$ as the 
classification subgroup, as suggested by the Hamiltonian
approach of Ref.(6). This alternative choice of the structure 
group automatically leads to the decomposition of the 
fourvector--valued coframe studied above into an $SO(3)$ triplet
plus an $SO(3)$ singlet respectively, the singlet characterizing
the time component of the coframe. The invariance of the time 
component of the coframe under $SO(3)$ transformations means in 
fact that it is Poincar\'e invariant. 

Let us at the first place decompose the Lorentz generators into 
boosts $K_a$ and space rotations $S_a\,$, respectively defined as 
$$K_a :=\,2\,L_{a0}\quad\,,\quad S_a 
:=-\epsilon _a{}^{bc} L_{bc}\qquad\qquad (a=\,1\,,2\,,3)
\,.\eqno(\z)$$
Their commutation relations are given in (A.5). The infinitesimal
group elements of the whole Poincar\'e group become parametrized as  
$$g=\,e^{i\,\epsilon^\alpha P_\alpha }
e^{i\,\beta ^{\alpha\beta}L_{\alpha\beta}} 
\approx\,1+i\,\left(\epsilon ^0 P_0 +\epsilon ^a P_a +\xi ^a K_a
+\theta ^a S_a\,\right)\,.\eqno(\z)$$
The difference with respect to the previous nonlinear
realization given by the choice (4.2) consists in the distinct
canonical projection we define in the group space. In other
words, we now choose $SO(3)$ as the structure group of the 
Poincar\'e principal fibre bundle. Accordingly, the 
parametrization of the fibres and the sections is no more as in
(4.2), but the following. The (infinitesimal) group elements of the
structure group $SO(3)$ are taken to be 
$$h =e^{i\,{\bf{\Theta}} ^a S_a}\approx\,1
+i\,{\bf{\Theta}} ^a S_a \,,\eqno(\z)$$
and on the other hand 
$$\sigma =e^{-i\,x ^\alpha P_\alpha}e^{i\,\lambda ^a K_a}\,,\eqno(\z)$$
where $x ^\alpha \,$ and $\lambda ^a $ are the (finite) coset
parameters. 

According to (2.8) {\it cum} (4.8--10), the variation of the 
translational parameters reads
$$\eqalign{\delta x^0=&-\xi ^a x_a-\epsilon ^0\,,\cr 
\delta x ^a =&\,\,\epsilon ^a{}_{bc}\theta ^b x^c -\xi ^a x^0 
-\epsilon ^a\,,\cr }\eqno(\z)$$
which coincides exactly with (4.3) since $\xi ^a :=\beta ^{a0}$ and
$\theta ^a :=-{1\over 2}\epsilon ^a{}_{bc}\,\beta ^{bc}\,$, as
read out from (4.8). Thus, the translational parameters still play 
the role of coordinates. In addition, we obtain the variations
of the boost parameters of (4.10) as 
$$\delta\lambda ^a =\,\epsilon ^a {}_{bc}\theta ^b \lambda ^c 
+\xi ^a |\lambda |\coth |\lambda | +{{\lambda ^a\lambda _b\xi
^b}\over{|\lambda |^2}}\left( 1-|\lambda |\coth |\lambda |\,\right)
\,,\eqno(\z)$$
being
$$|\lambda |:=\,\sqrt{\lambda _1{}^2 +\lambda _2{}^2 +\lambda _3{}^2 }
\,.\eqno(\z)$$
The meaning of $\lambda ^a$ will be discussed later. On the
other hand, according to (2.7), the infinitesimal action of the 
Poincar\'e group on arbitrary fields $\psi$ of a given
representation space of the $SO(3)$ group reads 
$$\delta\psi =\,i\, {\bf{\Theta}} ^a\rho\left( S_a\,\right)\psi
\,,\eqno(\z)$$
being $\rho\left( S_a\,\right)$ an arbitrary representation of 
$SO(3)\,$, and ${\bf{\Theta}} ^a$ the nonlinear $SO(3)$
parameter in (4.9), calculated from (2.8) to be 
$${\bf{\Theta}} ^a =\,\theta ^a +\epsilon ^a{}_{bc}
{{\lambda ^b \xi ^c}\over{|\lambda |}}\tanh\left( {{|\lambda |}
\over 2}\right) \,.\eqno(\z)$$

Let us now introduce the suitable gauge fields. In terms of the 
ordinary linear Poincar\'e connection (4.4), which may be
rewritten as 
$$\Omega :=-i\,{\buildrel (T)\over{\Gamma ^\alpha}} P_\alpha 
           -i\,\Omega ^{\alpha\beta}L_{\alpha\beta} 
          =-i\,{\buildrel (T)\over{\Gamma ^0}} P_0 
           -i\,{\buildrel (T)\over{\Gamma ^a}} P_a 
           +i\,{\buildrel (K)\over{\Gamma ^a}} K_a 
           +i\,{\buildrel (S)\over{\Gamma ^a}} S_a \,,\eqno(\z)$$
we define the nonlinear connection (2.9) as
$$\Gamma :=\,\sigma ^{-1}\left(d\,+\Omega\,\right) \sigma 
          =-i\,\vartheta ^\alpha P_\alpha 
           -i\,\Gamma ^{\alpha\beta}L_{\alpha\beta}
          =-i\,\vartheta ^0 P_0 
           -i\,\vartheta ^a P_a 
           +i\,X^a K_a 
           +i\,A^a S_a \,.\eqno(\z)$$
The translational nonlinear connections $\vartheta ^0\,$, 
$\vartheta ^a$ in (4.17) are the coframe components, 
whereas the vector--valued 1--forms $X^a \equiv\,\Gamma ^{0a}$ 
represent the gauge fields associated to the boosts; all of them
vary as $SO(3)$ tensors. Only $A^a\equiv\,{1\over2}\,
\epsilon ^a{}_{bc}\,\Gamma ^{bc}$ behaves as an ordinary
rotational connection. In fact, making use of (2.11) we find  
$$\eqalign{\delta\vartheta ^0 &=\,0\cr 
\delta\vartheta ^a &=\,\epsilon ^a{}_{bc}\,{\bf{\Theta}} ^b\,
\vartheta ^c \cr 
\delta X^a &=\,\epsilon ^a{}_{bc}\,{\bf{\Theta}} ^b\,X^c \cr 
\delta A^a  &=-D\,{\bf{\Theta}} ^a :=-\left( d\,{\bf{\Theta}} ^a 
+\epsilon ^a{}_{bc}\,A^b\,{\bf{\Theta}} ^c\,\right)\,.\cr}\eqno(\z)$$
In addition, the trivial metric $\delta _{ab}$ is a natural 
$SO(3)$ invariant.

As we have repeatedly pointed out before, the time component 
$\vartheta ^0$ of the coframe is invariant under local Poincar\'e 
transformations. Let us see how it happens. Making use of the
four--dimensional representation (A.6,7) of the Lorentz group, the 
relation between the nonlinear coframe components in (4.17) and the
Lorentz covector valued coframe in (4.6) may be expressed in the
simple form 
$$\pmatrix{\vartheta ^0\cr \vartheta ^a}=
\,e^{-i\,\lambda ^a \rho\left(K_a\,\right)}
\pmatrix{\tilde{\vartheta }^0\cr \tilde{\vartheta }^a}\,,\eqno(\z)$$
or, more explicitely
$$\eqalign{\vartheta ^0&=\,\tilde{\vartheta }^0\cosh |\lambda | 
+\tilde{\vartheta }^a \,{{\lambda _a}\over{|\lambda |}}\sinh
|\lambda |\,,\cr
\vartheta ^a&=\,\tilde{\vartheta }^a +\tilde{\vartheta }^b 
\,{{\lambda _b\lambda ^a}\over{|\lambda |^2}}\left(\cosh 
|\lambda |-1\right) +\tilde{\vartheta }^0\, {{\lambda ^a}
\over{|\lambda |}}\sinh |\lambda |\,.\cr }\eqno(\z)$$
The matrix $e^{-i\,\lambda ^a \rho\left(K_a\,\right)}\,$, see
(A.7), performs a change of basis leading from the Lorentz 
covector--valued 1--forms in the r.h.s. of (4.19), varying linearly as 
$$\delta\,\pmatrix{\tilde{\vartheta }^0\cr \tilde{\vartheta }^a}
=\,i\,\left[ \xi ^a \rho\left( K_a\,\right)
+\theta ^a \rho\left( S_a\,\right)\,\right]\,
\pmatrix{\tilde{\vartheta }^0\cr \tilde{\vartheta }^a}\,,\eqno(\z)$$
to the $SO(3)$ quantities in the l.h.s., whose variations are 
specified in (4.18). In fact, taking into account the transformation
properties of the coset parameter $\lambda ^a$ as given by
(4.12), it is easy to verify how the nonlinear realization splits
the four--dimensional representation into the SO(3) singlet 
$\vartheta ^0$ plus the SO(3) triplet $\vartheta ^a$ respectively. 
\bigskip\bigskip

\sectio{\bf Poincar\'e invariant spacetime foliation.}
\bigskip
The essential topological features of time are continuity
and one--dimensionality. We will not discuss the former, which
defines the vicinity of time instants in an obvious way, and
neither the possible global topology of time, homeomorhic to the 
real line. We will be exclusively concerned with the problem of 
defining a one--dimensional time direction inside the original
four--dimensional spacetime manifold. The possibility of
recovering a suitable notion of time, which reduces to the usual
one of Special Relativity in the absence of Gravity, should be a
consequence of a particular foliation of the spacetime, in such
a way that the resulting foliation direction makes physical
sense and possesses several properties one expects from the time.
In other words, according to Frobenius' theorem$^{(13)}$, one
has to identify a certain 1--form, say $u\,$, such that it
satisfies the foliation condition 
$$u\wedge d\,u =\,0\,.\eqno(\z)$$ 
>From (5.1) follows that $u=Ndt\,$, being $N$ the lapse function and $t\,$ 
the time parameter. The dual vector to $u\,$, denoted by $n\,$,
is defined from $u$ by means of the relation $n\rfloor u=1\,$. It
is a timelike vector field with the structure $n={1\over N}
\left(\partial _t -N^A\partial _A\,\right)\,$, where $N^A$
stands for the shift functions and $\partial _A$ represent the
derivatives with respect to the spatial coordinates. The time
vector $n$ then defines a preferred orientation in the
underlying manifold. Every value of the time parameter $t\,$
determines a spatial hypersurface which does not intersect
the hypersurfaces corresponding to distinct time values, i.e. 
the spacetime is foliated into spatial sheets. Any arbitrry
p--form $\alpha $ may thus be decomposed into a longitudinal 
part along the time direction plus a transversal part orthogonal
to it as $\alpha = u\wedge\left( n\rfloor\alpha\,\right)+n\rfloor
\left( u\wedge\alpha\,\right)\,$. When referred to the spatial
hypersurfaces, we will call $\alpha _{\bot}:= n\rfloor\alpha$ 
the normal part and $\underline{\alpha }:=n\rfloor\left( 
u\wedge\alpha\,\right)$ the tangential part.

The foliability of the spacetime manifold is the fundamental
requirement for a physical time to be well defined in the
context of General Relativity or of other possible approaches 
to the theoretical description of Gravity. But this condition is
by no means sufficient to uniquely characterize the time. In 
principle, the foliation performed with respect to the time
direction $u$ is merely topological, and thus extrinsic to the 
dynamical aspects represented by the gauge fields. The 1--form 
$u$ satisfying the Frobenius' condition (5.1) determines a
topological property of the underlying manifold, but it is 
introduced by hand without being {\it a priori} identifyable
with any dynamical object. One would expect, in a dynamical
theory of spacetime, that time should be related to the metric
tensor or to the {\it vierbeine}. In fact, let us consider the
coframe (4.6) with linear Lorentz indices, or alternatively let us 
define, in the context of ordinary General Relativity, a 
{\it vierbein} $e^\alpha{}_i$ which solders the manifold to its 
tangential spaces at any point. The tangential spaces are
Minkowskian, since Special Relativity is suposed to hold for
locally inertial reference frames. Thus, the coframes defined as
$\tilde{\vartheta}^\alpha := e^\alpha{}_i\,d x^i$ behave locally 
as Lorentz covectors. The following discussion holds for both
approaches. 

The time component $\tilde{\vartheta}^0$ of the coframe
introduced above, what we will call the dynamical time,
decomposes with respect to the topological time direction $u$ as
$\tilde{\vartheta}^0=u\wedge\tilde{\vartheta}^0_{\bot} 
+\underline{\tilde{\vartheta}}^0\,$, with a nonvanishing
contribution transversal to $u\,$. A seemengly natural choice to
define a unique physical time consists in aligneing both, the
topological and the dynamical time directions, by requiring 
$\underline{\tilde{\vartheta}}^0 =0\,$, or equivalently 
$$u=\tilde{\vartheta}^0\,.\eqno(\z)$$ 
The resulting coframe adapted foliation corresponds in fact to
the so called {\it time gauge} introduced by Schwinger in the 
literature$^{(14)}$. Unfortunately, the assumption (5.2) breaks the
local Lorentz symmetry of the theory. In fact, the foliation 
condition (5.1) involves not a covariant but an ordinary
differential, so that it only remains an invariant condition if
$u$ itself is invariant, which is not the case as far as 
$\tilde{\vartheta}^0$ transforms as the time component of a
four--covector. Thus, apparently, the price one has to
pay to define a single physical time in the presence of Gravity 
is that one has to fix the {\it time gauge}, loosing the local 
covariance under Lorentz transformations.

Before presenting our own solution to this problem, let us
summarize the desirable features one whishes to require from 
time. Fundamentally, one should identify a certain 1--form 
suitable to define a topological time direction on the
underlying fourdimensional coordinate manifold, i.e. a 1--form 
$u$ on which one could impose the Frobenius' foliation condition
(5.1). Furthermore, the candidate to induce the spacetime foliation
should preferably have the meaning of the dynamical time
component $\tilde{\vartheta}^0$ of a coframe, as in (5.2), in order
to define a single time with the topological and the dynamical 
time directions aligned, and thus interpretable as the unique 
physical time. On the other hand, if possible, one would expect 
to perform the foliation without breaking the gauge symmetry. 

As discussed above, there exist neither absolute rest nor
absolute motion on a topological spacetime manifold. Both, the 
spatial positions and motions are relative to physical
references. This assert holds also for the time evolution. It 
cannot be merely characterized topologically, since the
topological time is not directly observable. A physical
evolution process has to be necessarily evaluated with respect
to a physical {\it clock}, which allows to measure the relative 
rate of change. In our proposal, the time evolution will be 
referred to the natural time coframe $\vartheta ^0$ in (4.17--20). 
The foliation condition becomes expressable in terms of dynamical 
objects, notwithstanding its topological nature, and its meaning
and dynamical implications clarify the role played by time in 
Physics. The existence of the invariant time component of the 
coframe, see (4.18), enables us to perform an invariant foliation 
adapted to the nonlinear realization of the Poincar\'e group of 
previous section. The Frobenius' foliation condition (5.1) takes
the form 
$$\vartheta ^0\wedge d\,\vartheta ^0 =\,0\,.\eqno(\z)$$
In view of (4.18a), eq.(5.3) is Poincar\'e invariant, thus defining an
invariant foliation. Eq.(5.3) constitutes the integrability
condition for $\vartheta ^0\,$. From it follows 
$$\vartheta ^0 =\,u^0\,d\,\tau\,,\eqno(\z)$$
with $\tau$ as a dynamical time parameter. Observe that it is
absolutely different from the time coordinate. From the 1--form 
basis (4.19), we define its dual vector basis $e_\alpha $ such that 
$e_\alpha\rfloor\vartheta ^\beta =\,\delta _\alpha ^\beta\,$,
and we identify $e_{_0}$ as the invariant timelike vector field 
along which the foliation of the spacetime is defined. The Lie 
derivative of any arbitrary p--form $\alpha$ with respect to 
$e_{_0}\,$ reads  
$${\it{l}}_{e_{_0}}\alpha :=
\,d\,\left( e_{_0}\rfloor\alpha\,\right) 
+ \left( e_{_0}\rfloor d\,\alpha\,\right) \,,\eqno(\z)$$
representing the time evolution of $\alpha\,$. We remark that
this evolution is not merely topological, but dynamical --being
$\vartheta ^0$ a gauge field-- and with well defined time
metricity, since $\vartheta ^0$ and thus $e_{_0}$ are invariant.
This means that, on very general dynamical grounds, we have 
identified a physical {\it clock time} $\vartheta ^0\,$, that is
a dynamical field with respect to which the time evolution of any
system makes sense. It is meaningless to conceive $\vartheta ^0$
as {\it flowing} itself. The "transcurse of time" is measured by
the rate of change of any other field with respect to it. In 
particular, $e_{_0}\rfloor\vartheta ^0=1$ is the generalization
of the fact that $dt/dt=1$ in Newtonian mechanics, so that the
rate of change of time relative to itself is trivially constant
and positive.  

Further, $\alpha$ admits a decomposition into a longitudinal and
a transversal part with respect to the invariant vector field 
$e_{_0}\,$, namely
$$\alpha =\,\vartheta ^0\wedge\alpha _{\bot }
+\underline{\alpha }\,,\eqno(\z)$$
with $\alpha _{\bot}$ and $\underline{\alpha }$ respectively
defined as 
$$\alpha _{\bot}:=\,e_{_0}\rfloor\alpha\quad\,,\qquad 
\underline{\alpha }:=\,e_{_0}\rfloor\left(\, 
\vartheta ^0\wedge\alpha\,\right) \,.\eqno(\z)$$
Taking (5.5,6) into account, the decomposition the exterior
differential of $\alpha$ reads 
$$d\,\alpha =\,\vartheta ^0\wedge\left[\,{\it{l}}_{e_{_0}}
\underline{\alpha}-{1\over{u^0}}\,\underline{d}\,
\left( u^0\,\alpha _{\bot }\right)\,\right] +\underline{d}\,
\underline{\alpha}\,.\eqno(\z)$$
The invariance under time reparametrizations in the Hamiltonian
approach is related to the arbitrariness in the definition (5.7) of 
the components of the longitudinal parts of differential forms. 
But we will not develop this point here.

The differential forms may describe spatial motions, but they
are in fact relative ones, regulated by a certain gauge symmetry
group. The physically meaningful relative behavior is expressed
in terms of the coframes $\vartheta ^\alpha\,$. These
(nonlinear) gauge fields are 1--forms, i.e. intrinsic
objects not attached to absolute points. Their dual vector 
fields $e_\alpha$ represent the reference frames, and the 
relative four--velocity $\tilde{u}^\alpha$ of a different 
coframe $\tilde{\vartheta}^\alpha$ with respect to 
$\vartheta ^\alpha$ may be intrinsically defined to be 
proportional to $e_{_0}\rfloor\tilde{\vartheta}^\alpha\,$. Thus,
the coframes give account of both, the relative positions and
velocities. 

Instead, as discussed above, the coordinates are mathematical 
artefacts which in general do not posses any objective meaning.
In a covariant formulation, coordinate differences cannot be 
directly measured with the unit length. Nevertheless, under
certain assumptions, in particular in the absence of 
gravitational effects, the differentials of the coordinates 
become identifyable with the Lorentz linear coframes (4.6a)
themselves as 
$$\tilde{\vartheta}^\alpha =\delta ^\alpha _i\,dx^i
\,,\eqno(\z)$$ 
so that they coincide with dynamical quantities. The 
trivialization (5.9) of the coframes yields a correspondence 
between coordinates and measurable quantities, as in Newtonian
mechanics. However, considered from our point of view, also in
this case do the coordinates be of gauge theoretical origin,
since they still are parameters of the base space $G/H$ of 
the dynamical gauge theory of spacetime. Thus, even the 
special--relativistic kinematics is inseparable from the
spacetime dynamics. We can interpret the particular case (5.9) as 
the origin of the coordinates of Special Relativity, holding
when Gravity is negligible. Moreover, let us see how, in the 
absence of Gravitation, the invariant time (4.20a) reduces to the 
proper time of Special Relativity. According to (5.9), we identify 
$\tilde{\vartheta}^0 =dx^0 =:c\,dt\,,\quad\tilde{\vartheta}^a 
=\,dx^a\,$. Substituting these values in (4.20a), from Fermat's 
principle $\delta\int\vartheta ^0 =\,0$ we get 
$$\lambda ^a=-{{v^a}\over{|v\,|}}{\rm arctanh}\left({v\over c}
\right)\,.\eqno(\z)$$
Taking this value for $\lambda ^a\,$, the invariant time 
component of the coframe in (4.20) reduces to 
$$\vartheta ^0 = c\,dt\,\sqrt{1-v^2/c^2}\,,\eqno(\z)$$
that is, $c$ times the proper time, and $\vartheta ^a$ vanishes.
\bigskip\bigskip 

\sectio{\bf Nonlinear realizations and unitary gauge}
\bigskip 
The physically relevant fields of a dynamical theory do in
general not coincide with the original degrees of freedom
present in the action. To calculate the number of dynamical
fields, one has to substract the number of constraints plus 
the order of the symmetry group involved. Thus, in a gauge
theory it is necessary to identify the complete set of
constraints and fix the gauge, in order to deal only with 
physical degrees of freedom. The interested reader can find a
detailed derivation of the former for the Poincar\'e Gauge Theory 
in Ref.(6). With respect to the gauge fixing, at least two 
different approaches are possible. At the first place, the 
ordinary gauge fixing procedure consists in establishing 
conditions on the fields, breaking the gauge symmetry. We will 
not enter technical details, but we mention that the gauge
fixing should be performed after renormalization, since several 
critical phenomena are associated to the propagation of 
non--physical degrees of freedom. The second method we want to
mention is the unitary gauge fixing.

The unitary gauge procedure makes use of the symmetry properties
to covariantly eliminate the non--physical degrees of freedom of
the theory. The fields eliminable by means of a suitable
symmetry transformation are the Goldstone bosons, which are 
isomorphic to the group parameters. They are non--physical, and 
they become gauged away, embedded in the remaining fields of the
theory. We point out the similitude between the absorption of
the Goldstone fields in the unitary gauge mechanism, and that of
the coset parameters of the nonlinear realizations, which do not
explicitely appear in the theory since they are embedded in the
nonlinear fields. In fact, the nonlinear approach provides the 
natural language to deal with the unitary gauge procedure. Let 
us illustrate the relation between both by examining the unitary
gauge, as it commonly appears in the standard model.
Accordingly, we breafly outline the nonlinear gauge approach to 
$SU(2)\otimes U(1)\,$. 

The generators $T_a\,,Y\,$, of $SU(2)$ and $U(1)$ respectively, 
satisfy the standard commutation relations, and the linear 
connection of the group reads 
$$\Omega :=-ig\,A^a T_a -i{{\,\,g\,'}\over 2}\,B Y\,.\eqno(\z)$$
For further convenience we redefine the generators as 
$$T^{\,+}:=\,{1\over{\sqrt{2}}}\left( T_1+iT_2\,\right)\,,\qquad
T^{\,-}:=\,{1\over{\sqrt{2}}}\left( T_1-iT_2\,\right)\,,\qquad 
Q:=\,T_3+{Y\over 2}\,.\eqno(\z)$$
The generators $Q$ are those of the electromagnetic $U(1)_{el}$
group. We will perform the nonlinear realization with this group
as the structure group. Thus, we apply the general formula (2.8)
with the particular choices
$$g=\,e^{\,i\,\left( \epsilon ^{\,+}T^{\,+} 
+\epsilon ^{\,-}T^{\,-} +\epsilon {\,^0}T_3+y\,Q\,\right)}
\,,\qquad h =e^{i\,\lambda Q}\,,\qquad 
\sigma =\,e^{i\,\lambda ^{\,+}T^{\,+}}e^{i\,\lambda ^{\,-}T^{\,-}}
e^{i\,\lambda ^{\,0}T_3}\,.\eqno(\z)$$
The action (2.7) of the whole group on arbitrary fields $\psi$ of 
a given representation space of the classification subgroup 
$U(1)_{el}$ reads infinitesimally
$$\delta\psi =\,i\,\lambda\rho\left( Q\,\right)\psi\,,\eqno(\z)$$
being $\lambda $ the nonlinear $U(1)_{el}$ parameter, and 
$\rho\left( Q\,\right)$ a suitable representation of the
$U(1)_{el}$ group. Let us show how the fields $\psi$ in (6.4),
characteristic for the nonlinear approach, relate to the 
standard fields, say $\phi\,$, of the linear $SU(2)\otimes U(1)$
theory. We take in particular $\phi$ to be a complex doublet
such that $\phi ^{\dagger}{}\phi =\chi {}^2\,$. Its four degrees
of freedom can be rearranged as follows. Let us take the 
corresponding $2\times 2$ representation $T_3 ={1\over2}
\sigma _3 \,,Y=I\,$. According to (6.2c) we get 
$Q=\pmatrix{10\cr 00}\,$, and thus $\pmatrix{0\cr\chi}$ is an 
$U(1)_{el}$ scalar. Thus we can parametrize the linear field as
$$\phi =\,e^{i\,\xi ^{\,+}T^{\,+}}e^{i\,\xi ^{\,-}T^{\,-}}
e^{i\,\xi ^{\,0}T_3}\pmatrix{0\cr\chi}\,.\eqno(\z)$$
The reader will recognize in (6.5) the usual parametrization of the
Higgs multiplet with $\xi ^{\,+}\,,\xi ^{\,-}\,,\xi ^{\,0}$ 
as the Goldstone bosons. The formulation of the theory in the 
unitary gauge requires to perform a transformation with group 
parameters chosen to be functions of the fields of the theory in
such a way that they cancel out the physically superfluous
degrees of freedom. This is equivalent to realize the theory 
nonlinearly with suitably chosen field--dependent coset 
parameters. In fact, the nonlinear fields in (6.4) relate to the
linear ones as 
$$\psi =\sigma ^{-1}\phi\,,\eqno(\z)$$
with $\sigma ^{-1}$ the inverse of $\sigma$ in (6.3c). Thus, in the
present case it suffices to choose the coset parameters 
$\lambda ^{\,+}\,,\lambda ^{\,-}\,,\lambda ^{\,0}$ of
$\sigma$ to be respectively equal to the degrees of freedom 
$\xi ^{\,+}\,,\xi ^{\,-}\,,\xi ^{\,0}\,$, present in (6.5), to get 
$$\psi =\pmatrix{0\cr\chi}\eqno(\z)$$
in the unitary gauge. The non eliminable field $\chi$ in (6.7) is
the Higgs field, which remains as the only physical degree of 
freedom. Simultaneously, one has to transform the linear
connection (6.1) into 
$$\eqalign{\Gamma :&=\,\sigma ^{-1}\left(d\,+\Omega\,\right)\sigma\cr 
&=-ig\,\left( {\bf W}^{\,+}T^{\,+} 
+{\bf W}^{\,-}T^{\,-}\,\right) 
+{{ig}\over{\cos\theta _w}}\,{\bf Z}
\left( T_3 -\sin ^2\theta _w\,Q\,\right) 
-ie\,{\bf A}\,Q\,,\cr}\eqno(\z)$$
compare (6.8) with (2.9), being  
$$\theta _w:=\arctan {{\,\,g\,'}\over g}\quad\,,\qquad 
e:={{g\,g\,'}\over{\sqrt{g^2+g\,'^2}}}\,,\eqno(\z)$$
with the effective gauge fields suitably defined in terms of
the linear ones (6.1) and the Goldstone bosons. Since those are
equal to the coset parameters, they disappear as explicit
degrees of freedom, embedded in the redefined vector fields, 
whose variations read 
$$\delta\,{\bf W}^{\,+}
=\,i\,\lambda\,{\bf W}^{\,+}\,,\quad 
\delta\,{\bf W}^{\,-}=-i\,\lambda\,{\bf W}^{\,-}\,,\quad 
\delta\,{\bf Z}=\,0\,,\quad  
\delta\,{\bf A}=\,{1\over e}\,d\,\lambda\,.\eqno(\z)$$
The remarkable fact is that the unitary gauge procedure consists
in performing a particular transformation from the linear to a 
nonlinear realization of the gauge group. The tensorial
character of ${\bf W}^{\,+}$, ${\bf W}^{\,-}$ and ${\bf Z}$ is a
result of the nonlinear realization. We point out the analogy
between eqs. (6.6) and (4.19), to which we will return below.
The main difference resides in that the former relates a nonlinear
realization to the linear one, whereas the latter establishes a 
relation between two different nonlinear realizations, since the
translations are in both cases nonlinearly treated.

The unitary gauge fixing may be total or partial. The total 
one corresponds to the choice of the structure group to be 
$H=I\,$, which implies that all the group parameters are treated
as Goldstone fields, and subsequently supressed as dynamical
fields. On the other hand, the role of the partial unitary gauge
fixing is that of restricting the number of degrees of freedom
by eliminating those not corresponding to the structure group 
$H\,$. The connections and linear representations of $H$ remain 
unaltered. Since the gauge is fixed covariantly, the resulting 
nonlinear theory is formally identical to the linear one, but
being the Goldsone fields absent, it depends on a fewer number
of degrees of freedom. Those associated to the group parameters
of $H$ may be fixed by a subsequent symmetry breaking. As in the
ordinary gauge fixing, the unitary gauge is to be introduced
after the renormalization of the theory, in such a way that it
does not affect the quantization. Let us now apply the unitary
gauge fixing procedure to the gauge theory of spacetime outlined
above. 
\bigskip\bigskip 

\sectio{\bf The unitary gauge in Gravitation}
\bigskip 
The time component $\vartheta ^0$ of the coframe is trivially
longitudinal with respect to itself. Contrarily, 
$d\,\vartheta ^0$ presents in principle a nonvanishing 
contribution transversal to $\vartheta ^0\,$, such that in
general $\vartheta ^0\wedge d\,\vartheta ^0\neq 0\,$, in
disagreement with the integrability condition of $\vartheta ^0$
represented by the Frobenius foliation condition (5.3). Thus, 
the presence of the transversal degrees of freedom of 
$d\,\vartheta ^0$ in a dynamical theory would constitute a 
topological obstruction to the integrability of the evolution 
equations. The unitary gauge provides the method to eliminate the
transversal degrees of freedom of $d\,\vartheta ^0$ as Goldstone
bosons, and thus to guarantee both, the foliation of the
spacetime and the integrability of the dynamical equations. 

As pointed out in the previous section, in order to fix the unitary 
gauge we have to choose the group parameters involved in a
general nonlinear realization in such a way that they coincide
with suitable functions of the fields, capable to cancel them
out as non physical Goldstones. In the present case, the choice
of the group parameters will be somewhat more complicated as in 
the example considered above, where we simply took them to be
equal to the superfluous fields. Here we have to choose the
$\lambda ^a$'s in (4.19) to be certain functions of the Lorentz
linear coframes in the r.h.s. of (4.19) itself in order to fix 
the unitary gauge in such a way that it supresses the Goldstone 
fields associated to the boosts. We make use of a 1--form 
$\mu\,d\rho$ to express the dependence on the coframes 
$\tilde{\vartheta}^\alpha\,$, which are represented by their
dual vectors $\tilde{e}_\alpha$ acting on it. For further 
convenience, we use the notation $\mu\,d\rho\equiv
u^0\,d\tau\,$. As we will see below, this 1--form corresponds to
the non eliminable Higgs--like field leaved by the unitary gauge
fixing. At the first place, we define a three--velocity $v^a$
such that 
$${{v_a}\over{|v\,|}}:=-{{\left(\tilde{e}_a\rfloor u^0\,d\tau
\,\right) }\over{\sqrt{\left(\tilde{e}_{_0}\rfloor u^0\,d\tau
\,\right) ^2-1}}}\,.\eqno(\z)$$
Despite it depends on the four vector fields involved, it is
subjected to the constraint 
$$\left(\tilde{e}_{_0}\rfloor u^0\,d\tau\,\right) ^2 -
\left(\tilde{e}_a\rfloor u^0\,d\tau\,\right)
\left(\tilde{e}^a\rfloor u^0\,d\tau\,\right)=\,1\,,\eqno(\z)$$
so that only three are in fact relevant. Now we choose the boost
parameters $\lambda ^a$ in (4.19) to be 
$$\lambda ^a =-{{v^a}\over{|v\,|}}{\rm arctanh}\sqrt{
1-1/\left(\tilde{e}_{_0}\rfloor u^0\,d\tau\,\right) ^2}
\,.\eqno(\z)$$
This value of $\lambda ^a$ fixes the unitary gauge. Let us see
it in some detail. From (7.3) follows 
$$|\lambda | =\,{\rm arctanh}\sqrt{ 1-1/\left(\tilde{e}_{_0}
\rfloor u^0\,d\tau\,\right) ^2}\,,\eqno(\z)$$
which yields 
$$\cosh |\lambda |=\left(\tilde{e}_{_0}
\rfloor u^0\,d\tau\,\right)\,.\eqno(\z)$$ 
On the other hand, substituting (7.4) in (7.3) we get 
$${{\lambda ^a}\over{|\lambda |}}=-{{v^a}\over{|v\,|}}
\,.\eqno(\z)$$
Making then use of definition (7.1) {\it cum} (7.5), one obtains 
$${{\lambda ^a}\over{|\lambda |}}\sinh |\lambda | =
\left(\tilde{e}_a\rfloor u^0\,d\tau\,\right)\,.\eqno(\z)$$
Eqs.(7.5,7) as derived from (7.3) lead to the main result we were
looking for. In fact, substituting them in (4.20), one obtains 
$$\eqalign{\vartheta ^0&=\,\tilde{\vartheta }^\alpha 
\left(\tilde{e}_\alpha\rfloor u^0\,d\tau\,\right)
=\,u^0\,d\tau\,,\cr
\vartheta ^a&=\,\tilde{\vartheta }^a +\tilde{\vartheta }^b 
\,{{v_b v^a}\over{|v\,|^2}}\left[\left(\tilde{e}_{_0}\rfloor 
u^0\,d\tau\,\right) -1\right] +\tilde{\vartheta }^0\, 
\left(\tilde{e}^a\rfloor u^0\,d\tau\,\right)\,.\cr }
\eqno(\z)$$
The time component depends no more on four, but only on one
degree of freedom. The unitary gauge leaves a unique
Higgs--like time field, satisfying the Frobenius foliation
condition (5.3). The remaining three degrees of freedom, associated
to the boosts and corresponding to the transversal part of 
$d\,\vartheta ^0\,$, have been gauged away as Goldsone fields.
Moreover, since $\vartheta ^0 =\,u^0\,d\tau$ is invariant, being
the gauge fixing condition (7.3) equivalent to the Frobenius
invariant foliation condition (5.3), the right transformation 
properties of (7.3), namely (4.12), are guaranteed, as can be checked
by explicit calculation.

In a rigorous treatment of Gravity, the unitary gauge should be
fixed after renormalization --if renormalization is possible at
all. The reason is that the transversal degrees of freedom of 
$d\,\vartheta ^0\,$, eliminated as Goldstone bosons, could
propagate (like the Goldstones in the standard model), and thus
play a role in the quantum approach. 

In our repeatedly cited Ref.(6), we have developed a Hamiltonian
formalism adapted to the Poincar\'e Gauge Theory of Gravitation.
There, the Frobenius foliation condition was introduced by hand
without further justification. However, the discussion of the
present paper on this subject helps understanding the meaning of
such assumption. The Hamiltonian equations found by us$^{(6)}$
are the Einstein ones in the unitary gauge, where the Goldstone
fields associated to the boosts are absent. The spacetime is
foliated, and consequently the evolution equations are
integrable. In the original theory both, first class constraints 
corresponding to the symmetries and second class constraints,
are present. After totally fixing the gauge, i.e, once we break
the residual $SO(3)$ symmetry, only second class constraints
will remain. This essential fact should be taken into account in
order to quantize Gravitation.
\bigskip\bigskip 

\sectio{\bf Conclusions}
\bigskip 
We studied the gauge--theoretical foundations of the dynamical 
theory of spacetime. The Poincar\'e group $P$ realized nonlinearly
on its own parameter space, structurated as a principal fibre
bundle $P\left( P/SO(3)\,,SO(3)\right)\,$, is the unique axiomatic 
assumption we need to derive the main features of spacetime in 
relativistic Physics. The differentiable manifold, the dynamical
coframes and the Lorentz connections are all derived in a
deductive way. The special role played by the structure group
$SO(3)$ has to do with the existence of time. In fact, the
dynamical time is represented by the Poincar\'e invariant $SO(3)$ 
singlet $\vartheta ^0\,$. The unitary gauge fixing of the
boost symmetry, which cancels out the Goldsone fields
corresponding to the transversal degrees of freedom of 
$d\,\vartheta ^0\,$, gives rise to a foliation of spacetime
along the integrable $\vartheta ^0 =\,u^0\,d\tau$ direction. The
time evolution is defined as the Lie derivative along the
invariant time--like vector field $e_{_0}\,$, dual to 
$\vartheta ^0\,$. The Hamiltonian evolution equations of Gravity
derived by us in a previous paper are to be understood as the 
Einstein equations in the unitary gauge which guarantees their 
integrability.\bigskip\bigskip

\centerline {\bf Acknowledgements}\bigskip
We are very grateful to Dr. Juan Le\'on and Dr. J.F. Barbero for 
helpful discussions and constant interest in our work.
\bigskip\bigskip\bigskip

\centerline{\bf APPENDIX}
\bigskip
\noindent{\bf{B.--The Poincar\'e group in terms of boosts, rotations and
space and time translations}}\bigskip 
In the fourdimensional notation, the Lorentz generators 
$L_{\alpha\beta}$ and the translational generators
$P_\alpha\hskip0.3cm (\alpha\,,\beta =\,0...3)$ of the Poincar\'e
group satisfy the commutation relations
$$\eqalign{\left[L_{\alpha\beta }\,,L_{\mu\nu }\right]\hskip0.10cm&=
-i\,\left( o_{\alpha [\mu } L_{\nu ]\beta}   
         - o_{\beta [\mu }L_{\nu ]\alpha }\right)\,,\cr 
\left[L_{\alpha\beta }\,, P_\mu\,\right]\hskip0.10cm&=
\,\,\,\,i\,o_{\mu [\alpha }P_{\beta ]}\,,\cr 
\left[ P_\alpha\,, P_\beta\,\right]\hskip0.10cm&
=\,\,\,0\,.\cr}\eqno(A.1)$$
We choose the invariant metric tensor to be 
$$o_{\alpha\beta}:=\,diag(-\,+\,+\,+\,)\,,\eqno(A.2)$$
and we define 
$$S_a :=-\epsilon _a{}^{bc} L_{bc}\,,\eqno(A.3)$$
$$K_a :=\,2\,L_{a0}\,,\eqno(A.4)$$
with $a\,,b$ running from 1 to 3. The generators (A.3) are those
of the $SO(3)$ group, and (A.4) correspond to the boosts. In
terms of them, and taking (A.2) into account, the commutation 
relations (A.1) transform into 
$$\eqalign{\left[ S_a\,,S_b\,\right]&=-i\,\epsilon _{ab}{}^c S_c\,,\cr
\left[ K_a\,, K_b\,\right]&=\,\,\,\,i\,\epsilon _{ab}{}^c S_c\,,\cr
\left[ S_a\,,K_b\,\right]&=-i\,\epsilon _{ab}{}^c K_c\,,\cr
\left[ S_a\,,P_0\,\right]&=\,\,\,\,0\,,\cr
\left[ S_a\,,P_b\,\right]&=-i\,\epsilon _{ab}{}^c P_c\,,\cr
\left[ K_a\,,P_0\,\right]&=\,\,\,\,i\,P_a\,,\cr
\left[ K_a\,,P_b\,\right]&=\,\,\,\,i\,\delta _{ab}P_0\,,\cr
\left[ P_a\,,P_b\,\right]&=\,
\left[ P_a\,,P_0\,\right] =\,
\left[ P_0\,,P_0\,\right] =\,0\,.\cr}\eqno(A.5)$$
In terms of the 4--dimensional representation of the Lorentz 
generators given by 
$$\rho\left(S_1\,\right):=
-i\pmatrix{0\hskip0.2cm 0\hskip0.2cm 0\hskip0.4cm 0\cr 
           0\hskip0.2cm 0\hskip0.2cm 0\hskip0.4cm 0\cr 
           0\hskip0.2cm 0\hskip0.2cm 0-1\cr
           0\hskip0.2cm 0\hskip0.2cm 1\hskip0.4cm 0}
\quad
\rho\left(S_2\,\right):=
-i\pmatrix{0\hskip0.4cm 0\hskip0.2cm 0\hskip0.2cm 0\cr 
           0\hskip0.4cm 0\hskip0.2cm 0\hskip0.2cm 1\cr 
           0\hskip0.4cm 0\hskip0.2cm 0\hskip0.2cm 0\cr
           0-1\hskip0.2cm 0\hskip0.2cm 0}
\quad
\rho\left(S_3\,\right):=
-i\pmatrix{0\hskip0.2cm 0\hskip0.4cm 0\hskip0.2cm 0\cr 
           0\hskip0.2cm 0-1\hskip0.2cm 0\cr 
           0\hskip0.2cm 1\hskip0.4cm 0\hskip0.2cm 0\cr
           0\hskip0.2cm 0\hskip0.4cm 0\hskip0.2cm 0}$$

$$\rho\left(K_1\,\right):=
\,i\pmatrix{0\hskip0.2cm 1\hskip0.2cm 0\hskip0.2cm 0\cr 
           1\hskip0.2cm 0\hskip0.2cm 0\hskip0.2cm 0\cr 
           0\hskip0.2cm 0\hskip0.2cm 0\hskip0.2cm 0\cr
           0\hskip0.2cm 0\hskip0.2cm 0\hskip0.2cm 0}
\quad 
\rho\left(K_2\,\right):=
\,i\pmatrix{0\hskip0.2cm 0\hskip0.2cm 1\hskip0.2cm 0\cr 
           0\hskip0.2cm 0\hskip0.2cm 0\hskip0.2cm 0\cr 
           1\hskip0.2cm 0\hskip0.2cm 0\hskip0.2cm 0\cr
           0\hskip0.2cm 0\hskip0.2cm 0\hskip0.2cm 0}
\quad 
\rho\left(K_3\,\right):=
\,i\pmatrix{0\hskip0.2cm 0\hskip0.2cm 0\hskip0.2cm 1\cr 
           0\hskip0.2cm 0\hskip0.2cm 0\hskip0.2cm 0\cr 
           0\hskip0.2cm 0\hskip0.2cm 0\hskip0.2cm 0\cr
           1\hskip0.2cm 0\hskip0.2cm 0\hskip0.2cm 0}\,,\eqno(A.6)$$
we calculate the matrix 
$$\eqalign{&e^{-i\,\lambda ^a \rho\left(K_a\,\right)}=\,
1-i\,{{\lambda ^a}\over{|\lambda |}}\rho\left( K_a\,\right)
\sinh |\lambda | -{{\lambda ^a\lambda ^b}\over{|\lambda |^2}}
\rho\left(K_a\,\right)\rho\left(K_b\,\right)
\left(\cosh |\lambda | -1\right)\cr
&=\pmatrix{1\hskip0.2cm 0\hskip0.2cm 0\hskip0.2cm 0\cr 
         0\hskip0.2cm 1\hskip0.2cm 0\hskip0.2cm 0\cr 
         0\hskip0.2cm 0\hskip0.2cm 1\hskip0.2cm 0\cr
         0\hskip0.2cm 0\hskip0.2cm 0\hskip0.2cm 1}+
\pmatrix{0\hskip0.2cm \lambda_1\hskip0.2cm \lambda_2\hskip0.2cm \lambda_3\cr 
         \lambda_1\hskip0.2cm 0\hskip0.3cm 0\hskip0.3cm 0\cr 
         \lambda_2\hskip0.2cm 0\hskip0.3cm 0\hskip0.3cm 0\cr
         \lambda_3\hskip0.2cm 0\hskip0.3cm 0\hskip0.3cm 0}
{{\sinh |\lambda |}\over{|\lambda |}}+
\pmatrix{|\lambda |^2\hskip0.2cm 0\hskip0.8cm 0\hskip0.8cm 0\cr 
         0\hskip0.4cm \lambda _1^2\hskip0.2cm 
             \lambda _1\lambda _2\hskip0.2cm \lambda _1\lambda _3\cr 
         0\hskip0.2cm \lambda _1\lambda _2\hskip0.2cm 
             \lambda _2^2\hskip0.4cm \lambda _2\lambda _3\cr
         0\hskip0.2cm \lambda _1\lambda _3\hskip0.2cm 
             \lambda _2\lambda _3\hskip0.4cm \lambda _3^2}
{{\left(\cosh |\lambda | -1\right)}\over{|\lambda |^2}}
\,,\cr}\eqno(A.7)$$
wich is extensively usec in section 4.
\bigskip\vfill\eject 

\centerline{REFERENCES}
\vskip1.0cm
\noindent\item{[1]} C. Rovelli, {\it Phys. Rev.} {\bf D 42}
(1990) 2638, {\it Phys. Rev.} {\bf D 43} (1991) 442, 
{\it Nuov. Cim.} {\bf 110} (1995) 81, and references threin

\item\qquad C.J. Isham, {\it Canonical Quantum Gravity and the
Problem of Time}, in Recent Problems in Mathematical Physics,
Salamanca, June 15--27, 1992

\item\qquad A. Anderson, gr--qc/9507038, gr--qc/9507039

\item\qquad J.B. Hartle, gr--qc/9509037

\noindent\item{[2]} R. Utiyama,  {\it Phys. Rev.} {\bf 101} (1956) 1597

\item\qquad T. W. B. Kibble, {\it J. Math. Phys.} {\bf 2} (1961) 212

\item\qquad D. W. Sciama, {\it Rev. Mod. Phys.} {\bf 36} (1964) 463 and 1103

\item\qquad A. Trautman, in {\it Differential Geometry},
Symposia Mathematica Vol. 12 (Academic Press, London, 1973), p. 139

\item\qquad A. G. Agnese and P. Calvini, {\it Phys. Rev.} {\bf
D 12} (1975) 3800 and 3804 

\item\qquad F.W. Hehl, P. von der Heyde, G. D. Kerlick
and J. M. Nester, {\it Rev. Mod. Phys.} {\bf 48} (1976) 393

\item\qquad P. von der Heyde, {\it Phys. Lett.} {\bf 58 A} 
(1976) 141

\item\qquad F.W. Hehl, {\it Proc. of the 6th Course of the School of
Cosmology and Gravitation on Spin, Torsion, Rotation, and
Supergravity}, held at Erice, Italy, May 1979, P.G. Bergmann, V.
de Sabbata, eds. (Plenum, N.Y. i980) 5

\item\qquad E.A. Ivanov and J. Niederle, {\it Phys. Rev.} {\bf D25} 
(1982) 976 and 988

\item\qquad D. Ivanenko and G.A. Sardanashvily, {\it
Phys. Rep.} {\bf 94} (1983) 1

\item\qquad E. A. Lord, {\it J. Math. Phys.} {\bf 27} (1986) 2415 and 3051

\noindent\item{[3]} S. Coleman, J. Wess and B. Zumino, {\it Phys. Rev.}
{\bf 117} (1969) 2239 

\item\qquad C.G. Callan, S. Coleman, J. Wess and B. Zumino, {\it Phys. Rev.}
{\bf 117} (1969) 2247

\item\qquad S. Coleman, {\it Aspects of Symmetry}. Cambridge
University Press, Cambridge (1985)

\noindent\item{[4]} A.B. Borisov and I.V. Polubarinov, {\it Zh.
ksp. Theor. Fiz.} {\bf 48} (1965) 1625, and V. Ogievetsky and 
I. Polubarinov, {\it Ann. Phys.} (NY) {\bf 35} (1965) 167

\item\qquad A.B. Borisov and V.I. Ogievetskii, {\it Theor.
Mat. Fiz.} {\bf 21} (1974) 329

\item\qquad A. Salam and J. Strathdee, {\it Phys. Rev.}
{\bf 184} (1969) 1750 and 1760

\item\qquad C.J. Isham, A. Salam and J. Strathdee, {\it Ann. of Phys.} 
{\bf 62} (1971) 98

\item\qquad L.N. Chang and F. Mansouri, {\it Phys.
Lett.} {\bf 78 B} (1979) 274, and {\it Phys. Rev.} {\bf D 17} (1978) 3168

\item\qquad K.S. Stelle and P.C. West, {\it Phys. Rev.} 
{\bf D 21} (1980) 1466

\item\qquad E.A. Lord, {\it{Gen. Rel. Grav.}} {\bf 19} (1987) 983, and 
{\it J. Math. Phys.} {\bf 29} (1988) 258

\noindent\item{[5]} A. L\'opez--Pinto, A. Tiemblo and R.
Tresguerres, {\it Class. Quantum Grav.} {\bf 12} (1995) 1327 

\item\qquad J. Julve, A. L\'opez--Pinto, A. Tiemblo and R.
Tresguerres, {\it Nonlinear gauge realizations of spacetime
symmetries including translations}, to appear in G.R.G. (1996)

\noindent\item{[6]} A. L\'opez--Pinto, A. Tiemblo and R.
Tresguerres, gr--qc/9603023

\noindent\item{[7]} K. Hayashi and T. Nakano, {\it Prog. Theor. Phys} 
{\bf 38} (1967) 491

\item\qquad K. Hayashi and T. Shirafuji, {\it Prog. Theor. Phys} 
{\bf 64} (1980) 866 and {\bf 80} (1988) 711

\item\qquad G. Grignani and G. Nardelli, {\it Phys. Rev.} 
{\bf D 45} (1992) 2719

\item\qquad E. W. Mielke, J.D. McCrea, Y. Ne'eman and F.W. Hehl 
{\it Phys. Rev.} {\bf D 48} (1993) 673

\noindent\item{[8]} S. Kobayashi and K. Nomizu, {\it Foundations
of Differential Geometry}, Vol. I, Interscience Publ. N.Y. 1963

\noindent\item{[9]} M. Minkowski (1908), in {\it H.A. Lorentz,
A. Einstein und H. Minkowski, Das Relativit\"atsprinzip}, 
Teubner Verlag, Leipzig, Berlin. 1920

\noindent\item{[10]} M. Friedman, {\it Foundations of
Space--Time Theories}, Princeton University Press. 1983

\noindent\item{[11]} E. Cartan, {\it Sur les vari\'et\'es … connexion
affine et la th\'eorie de la relativit\'e g\'en\'eralis\'ee}, Ouvres
completes, Editions du C.N.R.S. (1984), Partie III. 1, pgs. 659
and 921

\noindent\item{[12]} F. W. Hehl, J. D. McCrea,  E. W. Mielke, and
Y. Ne'eman {\it Found. Phys.} {\bf 19} (1989) 1075

\item\qquad F.W. Hehl, J.D. McCrea, E.W. Mielke, and Y. Ne'eman, {\it
Physics Reports} {\bf 258} (1995) 1, and references therein

\noindent\item{[13]} F.W. Warner, {\it Foundations of 
differentiable manifolds and Lie groups}, {\it Scott, Foresman 
and Company}, Glenview, Illinois (1971)

\item\qquad R.P. Wallner, Ph.D. Thesis, University of
Vienna (1982), {\it Phys. Rev.} {\bf D 42} (1990) 441, 
{\it Jour. Math. Phys.} {\bf 36}(1995) 6937

\item\qquad E.W. Mielke, {\it Phys. Lett.} {\bf A 149}
(1990) 345, {\it Ann. Phys.} (N.Y.) {\bf 219} (1992) 78

\noindent\item{[14]} J. Schwinger, {\it Phys. Rev.} {\bf 130} 
(1963) 1253

\bye